%% file: main.tex
\documentclass[amsmath,amssymb,reprint,aps,pra,floatfix,superscriptaddress]{revtex4-1}
\usepackage{bm}
\usepackage{amsmath}
\usepackage{graphicx}
\graphicspath{ {./Figures/}, {./SM/} }
\usepackage{bm}
\usepackage{braket}
\usepackage{epstopdf}
\usepackage{bm}
\usepackage{bbm}
\usepackage{lineno}
\usepackage{hyperref}

\begin{document}

\title{Tuning between photon-number and quadrature measurements \\ with weak-field homodyne detection}

\author{G.S. Thekkadath}
\thanks{These two authors contributed equally.}
\affiliation{Clarendon Laboratory, University of Oxford, Parks Road, Oxford, OX1 3PU, UK}

\author{D.S. Phillips}
\thanks{These two authors contributed equally.}
\affiliation{Clarendon Laboratory, University of Oxford, Parks Road, Oxford, OX1 3PU, UK}

\author{J.F.F. Bulmer}
\affiliation{Clarendon Laboratory, University of Oxford, Parks Road, Oxford, OX1 3PU, UK}

\author{W.R. Clements}
\affiliation{Clarendon Laboratory, University of Oxford, Parks Road, Oxford, OX1 3PU, UK}

\author{A. Eckstein}
\affiliation{Clarendon Laboratory, University of Oxford, Parks Road, Oxford, OX1 3PU, UK}

\author{B.A. Bell}
\affiliation{Clarendon Laboratory, University of Oxford, Parks Road, Oxford, OX1 3PU, UK}

\author{J. Lugani}
\affiliation{Clarendon Laboratory, University of Oxford, Parks Road, Oxford, OX1 3PU, UK}

\author{T.A.W. Wolterink}
\affiliation{Clarendon Laboratory, University of Oxford, Parks Road, Oxford, OX1 3PU, UK}

\author{A. Lita}
\affiliation{National Institute of Standards and Technology, 325 Broadway, Boulder, CO 80305, USA}

\author{S.W. Nam}
\affiliation{National Institute of Standards and Technology, 325 Broadway, Boulder, CO 80305, USA}

\author{T. Gerrits}
\affiliation{National Institute of Standards and Technology, 325 Broadway, Boulder, CO 80305, USA}

\author{C.G. Wade}
\affiliation{Clarendon Laboratory, University of Oxford, Parks Road, Oxford, OX1 3PU, UK}

\author{I.A. Walmsley}
\affiliation{Clarendon Laboratory, University of Oxford, Parks Road, Oxford, OX1 3PU, UK}

\begin{abstract}
Variable measurement operators enable the optimization of strategies for testing quantum properties and the preparation of a range of quantum states.
Here, we experimentally implement a weak-field homodyne detector that can continuously tune between measuring photon numbers and field quadratures.
We combine a quantum signal with a coherent state on a balanced beam splitter and detect light at both output ports using photon-number-resolving transition edge sensors.
We observe that the discrete difference statistics converge to the quadrature distribution of the signal as we increase the coherent state amplitude.
Moreover, in a proof-of-principle demonstration of state engineering, we show the ability to control the photon-number distribution of a state that is heralded using our weak-field homodyne detector.
\end{abstract}

\maketitle

\textit{Introduction}. Counting particles is at the core of all measurements in quantum physics.
Many detection strategies used in, e.g., atomic~\cite{gerlach1922experimentelle}, nuclear~\cite{rutherford1908electrical}, or particle~\cite{glaser1952some} physics rely on this fundamental concept.
In optics, recent technological developments have enabled counting the number of quanta in light using photon-number-resolving detectors~\cite{H09}.
Since these detectors determine the exact number of photons in an optical signal, information about the complementary property, i.e., the wave nature of the signal, is lost.
In order to measure its wavelike properties, it is necessary to mix the signal with a phase-reference field.
This is usually achieved using homodyne detection, which uses a classical phase reference to measure the amplitude and phase of the signal~\cite{YS80}.
However, because the photodetectors used in arrangements like homodyne often have significant electronic noise, the phase reference must be quite strong, typically containing billions of photons~\cite{hansen2001ultrasensitive}.

Weak-field homodyne detection (WFHD) encompasses both the strategies mentioned above by first mixing a quantum signal with a weak phase reference and then detecting the resulting fields using photon-number-resolving detectors.
When the phase reference is comparable in strength to the signal, it cannot be treated as a classical field~\cite{B90,VG93,RCCBS95,BW97,TS04}.
In this regime, WFHD spans the region between the complementary detection strategies of direct photon counting and homodyne. 
This enables WFHD to reveal fundamental properties of light such as the presence of nonclassical correlations that would be hidden from conventional detection strategies~\cite{GPY88,BW98,KWM00}.

The versatility of WFHD also makes it a promising tool for optical quantum information processing.
On the one hand, homodyne detection performs phase-sensitive Gaussian measurements by projecting light onto continuous quadrature states.
This is the standard measurement used in quantum state tomography~\cite{lvovsky2009continous} and protocols that encode quantum information in continuous-variable degrees of freedom of light~\cite{braunstein2005quantum}.
On the other hand, photon counting performs phase-insensitive non-Gaussian measurements by projecting light onto discrete photon-number states.
This measurement is key when information is encoded in discrete variables (e.g., polarization, spatial mode) and also provides the non-Gaussian resource required for universal continuous-variable quantum computing~\cite{MvLGWRN06}. 
In principle, WFHD can be reconfigured to realize both of these measurements simply by controlling the strength of the phase reference.
Moreover, WFHD has the unique ability to perform phase-sensitive non-Gaussian measurements since it can access the region between homodyne and photon counting~\cite{PLBC-RSW09,ZC-RDPLJSPW12, DBJVDBW14}.
Such measurements provide a powerful state preparation and characterization tool, especially in hybrid discrete- and continuous-variable protocols~\cite{AN-NVF15}.
Although the tunability of WFHD has been investigated theoretically~\cite{B90,VG93,RCCBS95,BW97,TS04,PLBC-RSW09}, it has not yet been demonstrated experimentally due to limitations in the efficiency, noise, and dynamic range of the photon-number-resolving detectors used in previous implementations of WFHD~\cite{PLBC-RSW09, ZC-RDPLJSPW12, DBJVDBW14,ABOB17,OACPB18}.

In this paper, we overcome these limitations by using state-of-the-art transition edge sensors~\cite{LMN08}.
We experimentally demonstrate the ability to tune between photon-number and quadrature measurements of quantum signals.
We combine a heralded photon-number state (our signal) with a weak coherent state (our phase-reference) on a beam splitter. 
We then measure the photon-number difference between both output ports using transition edge sensors.
We observe that the discrete difference statistics converge to the quadrature distribution of the signal as we increase the coherent state strength.
Moreover, in a proof-of-principle demonstration of state engineering, we show the ability to control the photon-number distribution of a state that is heralded using WFHD.

\begin{figure}[t!]
	\includegraphics[width=0.55\columnwidth]{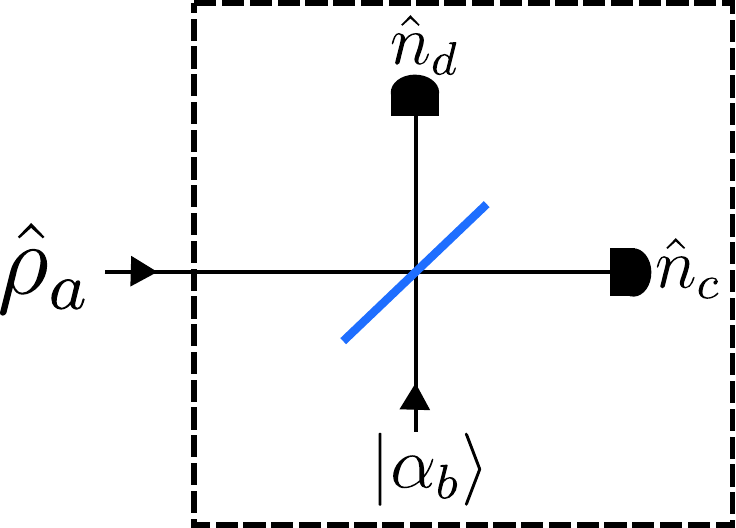}
	\caption{
	    \textbf{Weak-field homodyne detection.}
	     A signal $\hat{\rho}_a$ is combined with a coherent state $\ket{\alpha_b}$ on a balanced beam splitter.
	    The photon number of both outputs of the beam splitter is measured with photon-number-resolving detectors.
	    The dashed box constitutes a detector that can tune between projecting the input state $\hat{\rho}_a$ onto photon-number and quadrature states by varying the coherent state amplitude.
	}\label{fig:detector}
\end{figure}

\textit{Theory}. In homodyne detection, a signal $\hat{\rho}_a$ and coherent state of amplitude $\alpha$, $\ket{\alpha_b}$, are mixed on a balanced beam splitter (see Fig.~\ref{fig:detector}; subscripts label modes).
The difference in photon number $\Delta \hat{n} = \hat{n}_c - \hat{n}_d$ between both outputs of the beam splitter is given by
\begin{equation}
    \Delta \hat{n} = i(\hat{a}^\dagger \hat{b} - \hat{a}\hat{b}^\dagger),
    \label{eqn:photon_diff_op}
\end{equation}
where $\hat{a}$ and $\hat{b}$ are the input annihilation operators in modes $a$ and $b$, respectively. 
When $\alpha=0$, measuring the observable $\Delta \hat{n}$ with ideal number-resolving detectors projects $\hat{\rho}_a$ onto non-Gaussian photon-number states. 
In the limit of large $|\alpha|$, one can invoke the classical field approximation $\hat{b} \rightarrow |\alpha|e^{i\theta}$, where $|\alpha|$ and $\theta$ give the strength and phase of the coherent state, respectively~\cite{TS04}.
With this approximation, Eq.~\eqref{eqn:photon_diff_op} becomes:
\begin{equation}
    \label{eqn:photon_diff_op_classical}
    \Delta \hat{n}_{\mathrm{classical}} = i|\alpha|\hat{X}(\theta),
\end{equation}
where $\hat{X}(\theta) = \hat{a}^\dagger e^{i\theta} - \hat{a} e^{-i\theta}$ is the quadrature operator.
This is the usual treatment of homodyne detection found in textbooks~\cite{leonhardt1997measuring}.
In contrast to $\Delta \hat{n}$, measuring $\Delta \hat{n}_{\mathrm{classical}}$ projects $\hat{\rho}_a$ onto Gaussian quadrature states that are eigenstates of $\hat{X}(\theta)$~\cite{YS80}. 

The transition from $\Delta \hat{n}$ to $\Delta \hat{n}_{\mathrm{classical}}$ implies the ability to tune between photon-number and quadrature measurements by varying $|\alpha|$. 
To exploit this tunability in an experiment, one must be able to make $|\alpha|$ sufficiently large to ensure that the classical field approximation is valid. 
References~\cite{TS04,SDS17} estimate the required $|\alpha|$ to be $|\alpha| \gg N$, where $N$ is the average number of photons in $\hat{\rho}_a$.
This condition ensures that quantum fluctuations in the coherent state are larger than $N$.
If these fluctuations were smaller than $N$, i.e. $|\alpha| < N$, a measurement of $\Delta \hat{n}$ would reveal some information about $\hat{a}^\dagger\hat{a}$, the photon number of $\hat{\rho}_a$.
This information gained about $\hat{a}^\dagger\hat{a}$ comes at the cost of disturbing the measurement of $\hat{X}(\theta)$ since these two operators do not commute.

We investigate this transition in the particular case of a signal in a photon-number state, i.e. $\hat{\rho}_a = \ket{j_a}\bra{j_a}$~\cite{B90,VG93,RCCBS95}. 
Given some $|\alpha|$, the probability of measuring a difference of $\Delta n$ photons at the output of the beam splitter is given by (see the Supplemental Material~\cite{SM} for a derivation)
\begin{equation}
    \begin{split}
        P^{(j,\alpha)}(\Delta n) &= \sum_{m=\max(0,\Delta n)}^\infty \frac{e^{-|\alpha|^2}j!|\alpha|^{2(2m-\Delta n-j)}}{2^{2m-\Delta n}m!(m-\Delta n)!} \\
        &\times \left| \sum_{k=0}^j{m \choose m+k-j}{m-\Delta n \choose k}(-1)^{k}\right|^2.
    \end{split}
    \label{eqn:probabilityQuantum}
\end{equation}
Based on the arguments made above we expect that for $|\alpha| \gg j$, $P^{(j,\alpha)}(\Delta n)$ should converge to the measurement statistics of the observable $\Delta \hat{n}_{\mathrm{classical}}$, that is
\begin{equation}
    P_{\mathrm{classical}}^{(j,\alpha)}(\Delta n) = \frac{e^{-\Delta n^2/ 2|\alpha|^2}}{\sqrt{2\pi} 2^{j}j! |\alpha|}  \left | H_j\left( \Delta n/ \sqrt{2}\alpha \right) \right |^2,
    \label{eqn:probabilityClassical}
\end{equation}
where $H_j$ is the Hermite polynomial of degree $j$. Notably, $P_{\mathrm{classical}}^{(j,\alpha)}(\Delta n)$ is simply the quadrature distribution of a photon-number state $\ket{j}$ scaled by $|\alpha|$~\cite{leonhardt1997measuring}.
In the Supplemental Material~\cite{SM}, we generalize Eqs.~\eqref{eqn:probabilityQuantum} and~\eqref{eqn:probabilityClassical} to include imperfections that arise in our experiment such as optical losses, detection inefficiency, and mode mismatch between $\hat{\rho}_a$ and $\ket{\alpha_b}$.

\begin{figure}[t!]
	\includegraphics[width=1\columnwidth]{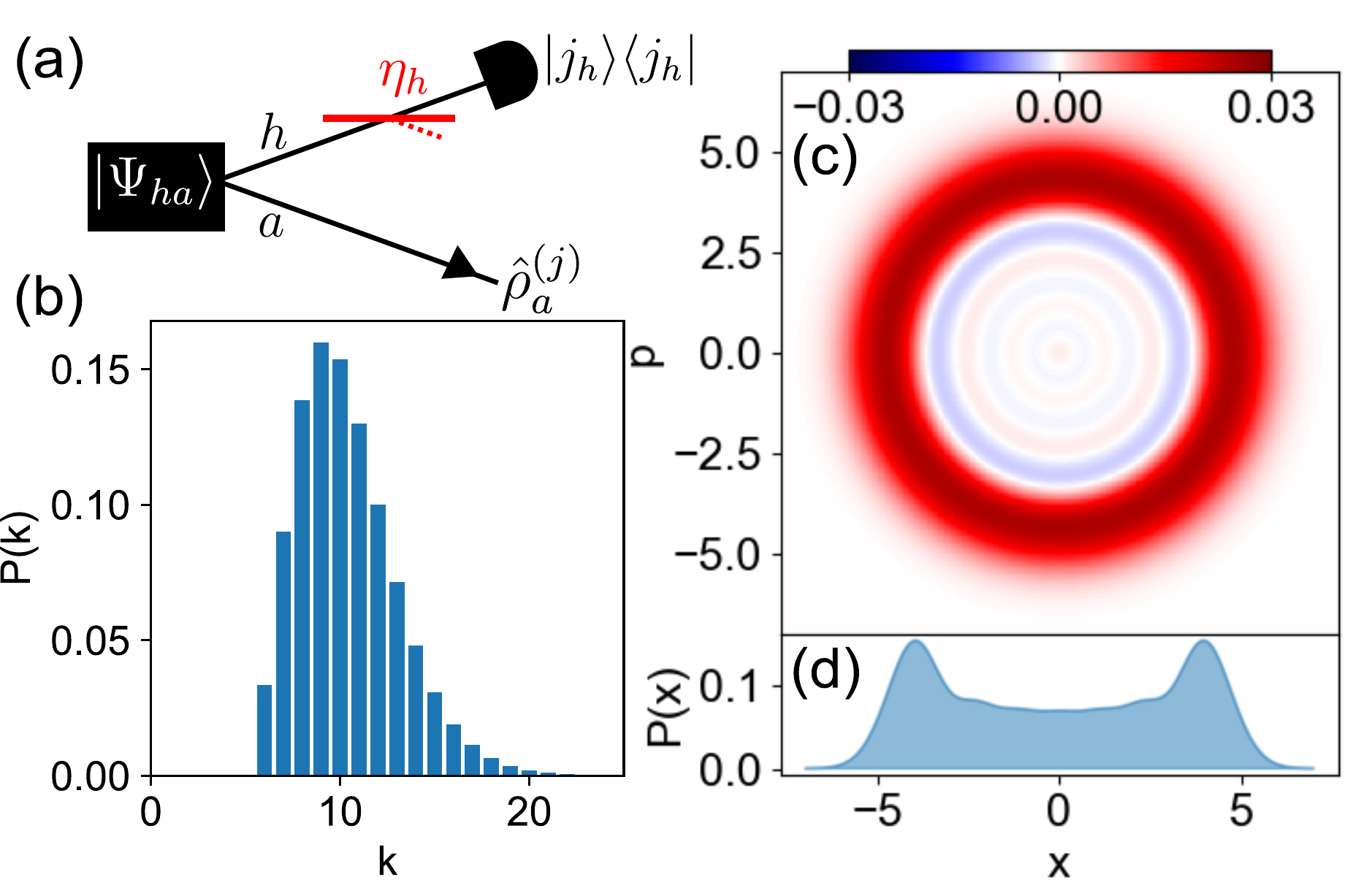}
	\caption{\textbf{Signal preparation.} 
	    (a) We prepare $\hat{\rho}^{(j)}_a$ by projecting mode $h$ of a two-mode squeezed vacuum state $\ket{\Psi_{ha}}$ onto a photon-number state $\ket{j_h}$ with efficiency $\eta_{h}$. In the remaining subplots, we consider the case of $j=6$ as an example.
	    (b) The simulated photon-number distribution $P(k) = \mathrm{Tr}_a(\ket{k_a}\bra{k_a}\hat{\rho}^{(6)}_a)$ with $N=10.4$.
	    (c) The simulated Wigner function $W(x,p) = \int_{-\infty}^{\infty} \braket{x+y|\hat{\rho}^{(6)}_a|x-y}\exp{\left(-2ipy\right)}dy$. The oscillations and negativity are caused by the abrupt cut in $P(k)$ below $k=6$.
	    (d) The simulated quadrature distribution $P(x) = \int_{-\infty}^{\infty} W(x,p)dp$, as would be measured by an ideal homodyne detector.
	}\label{fig:signal}
\end{figure}

\textit{Experimental overview}. Experimentally investigating the transition from $P^{(j,\alpha)}(\Delta n)$ to $P_{\mathrm{classical}}^{(j,\alpha)}(\Delta n)$ requires photon-number-resolving detectors with a large dynamic range.
The standard approach for photon-number-resolved measurements relies on multiplexing binary ``click" detectors~\cite{H09}.
However, this approach is not practical for our experiment since the required number of click detectors scales prohibitively with dynamic range~\cite{FJPF03}.  
Instead, we use superconducting transition edge sensors (TESs) which can resolve up to $\sim 20$ photons with $>$ 90\% efficiency~\cite{LMN08,HMGHLNNDKW15}.
TESs achieve photon-number resolution by acting as a bolometer, i.e. directly measuring the energy of the absorbed light.
More information on TES operation and readout is provided in the Supplemental Material~\cite{SM}.

To prepare our signal $\hat{\rho}_a$, we couple femtosecond laser pulses into a periodically-poled potassium titanyl phosphate (ppKTP) waveguide.
The waveguide produces an approximately spectrally decorrelated two-mode squeezed vacuum state,
\begin{equation}
\ket{\Psi_{ha}} = \sqrt{1-|\lambda|^2}\sum_{f=0}^{\infty}\lambda^f\ket{f_h,f_a},
\label{eqn:tmsv}
\end{equation}
via type-II parametric down conversion ($\lambda=\tanh{(r)}$ where $r$ is the squeezing parameter)~\cite{ECMS11}.
Mode $h$ is sent to a heralding TES detector.
By post-selecting on events where the heralding TES detects $j$ photons, we prepare our signal $\hat{\rho}^{(j)}_a$ in mode $a$ (see Fig.~\ref{fig:signal}).
Due to losses and imperfect detection efficiency ($\eta_h = 0.395 \pm 0.002$), we do not herald a pure photon-number state $\ket{j_a}$.
Rather, we herald a signal $\hat{\rho}^{(j)}_a$ which is a statistical mixture of photon-number states $\geq j$.
Despite the losses, $\hat{\rho}^{(j)}_a$ is still a nonclassical signal.
Namely, we measure that $\hat{\rho}^{(j)}_a$ has both sub-Poissonian~\cite{leonhardt1997measuring} and submultinomial~\cite{sperling2017detector} photon-number statistics when $j\geq 1$ (see the Supplemental Material~\cite{SM}).

\begin{figure}[t!]
    
    \includegraphics[width=1\columnwidth]{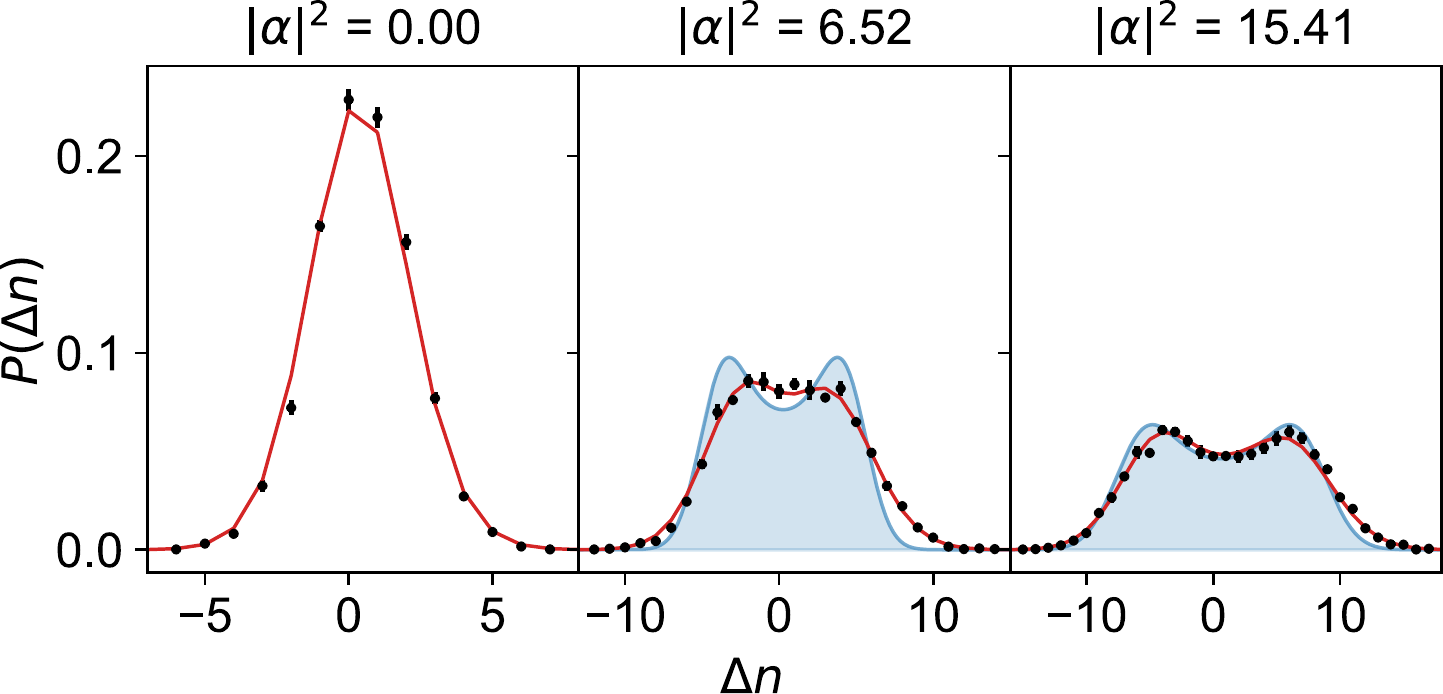}
	\caption{\textbf{Transition from a photon-number to a quadrature measurement.}
    	We plot the probability $P(\Delta n)$ to measure a photon-number difference $\Delta n$ with the signal $\hat{\rho}^{(6)}_a$.
    	As $|\alpha|^2$ increases, the agreement between the black data points and the expected quadrature distributions (blue regions) improves, indicating that our detector is performing a quadrature measurement.
        The red curves [blue regions] are calculated from $P^{(j,\alpha)}(\Delta n)$  [$P_{\mathrm{classical}}^{(j,\alpha)}(\Delta n)$], and include the effects of experimental imperfections.
        Error bars represent one standard deviation in 10 trials.
        Both the red and blue models are discrete; the lines interpolating between model points are merely to distinguish them from the data.
	}
	\label{fig:transitionPlot}
\end{figure}

To obtain a coherent state that is mode-matched to our signal, we prepare $\ket{\alpha_b}$ using a second ppKTP waveguide.
Unlike in the first waveguide, we stimulate the parametric down-conversion process by seeding it with light from a continuous-wave laser.
We use a strong seed to ensure that the spontaneous process is negligibly small relative to the stimulated process~\cite{aichele2002optical,LS13}.
As such, we generate a coherent state of light in the polarization mode orthogonal to the seed, $\ket{\alpha_b}$.
We then attenuate $\ket{\alpha_b}$ to the single-photon level with neutral density filters.
Since $\hat{\rho}^{(j)}_a$ has no defined phase relative to $\ket{\alpha_b}$, $P^{(j,\alpha)}(\Delta n)$ and $P_\mathrm{classical}^{(j,\alpha)}(\Delta n)$ do not depend on $\theta$ and hence we do not require control of $\theta$.

Both $\hat{\rho}^{(j)}_a$ and $\ket{\alpha_b}$ are coupled into fibers and temporally overlapped in a balanced fiber beam splitter using a delay stage.
The output modes of this beam splitter are then sent to two TESs. 
We measured the total system efficiency (i.e. coupling, transmission and detection efficiencies combined) in modes $c$ and $d$ to be $\eta_c = 0.274 \pm 0.001$ and $\eta_d = 0.354 \pm 0.002$, respectively.
Further details on the experimental setup can be found in the Supplemental Material~\cite{SM}.

\textit{Results}. In Fig.~\ref{fig:transitionPlot}, we show the measured photon-number difference statistics for $j=6$ and three different values of $|\alpha|^2$.
The measured statistics (black points) can be compared with the statistics expected from the theory model with and without the classical field approximation, i.e. $P_{\mathrm{classical}}^{(j,\alpha)}(\Delta n)$ (blue curves) and $P^{(j,\alpha)}(\Delta n)$ (red curves), respectively.
Both models include the effects of experimental imperfections such as detection inefficiencies and mode mismatch.
The parameters modeling these imperfections are determined from independent measurements (see the Supplemental Material~\cite{SM}).
For $|\alpha|^2 = 0$, $P^{(j,\alpha)}(\Delta n)$ is obtained by projecting $\hat{\rho}^{(6)}_a$ onto photon-number states.
In this case, $P_{\mathrm{classical}}^{(j,\alpha)}(\Delta n)$ is not defined.
For $|\alpha|^2 = 6.52$, the data points agree with $P^{(j,\alpha)}(\Delta n)$ but not with $P_{\mathrm{classical}}^{(j,\alpha)}(\Delta n)$, indicating that the classical field approximation is not yet valid.
However, for $|\alpha|^2=15.41$, there is good agreement between the data and both $P^{(j,\alpha)}(\Delta n)$ and $P_{\mathrm{classical}}^{(j,\alpha)}(\Delta n)$.
Surprisingly, it seems that the measurement is already projecting $\hat{\rho}^{(6)}_a$ onto quadrature states despite still being far from the regime where $|\alpha| \gg N$ (note that $N=10.4$ for $j=6$ due to loss in the herald mode).

\begin{figure}[t!]
    
    \includegraphics[width=1\columnwidth]{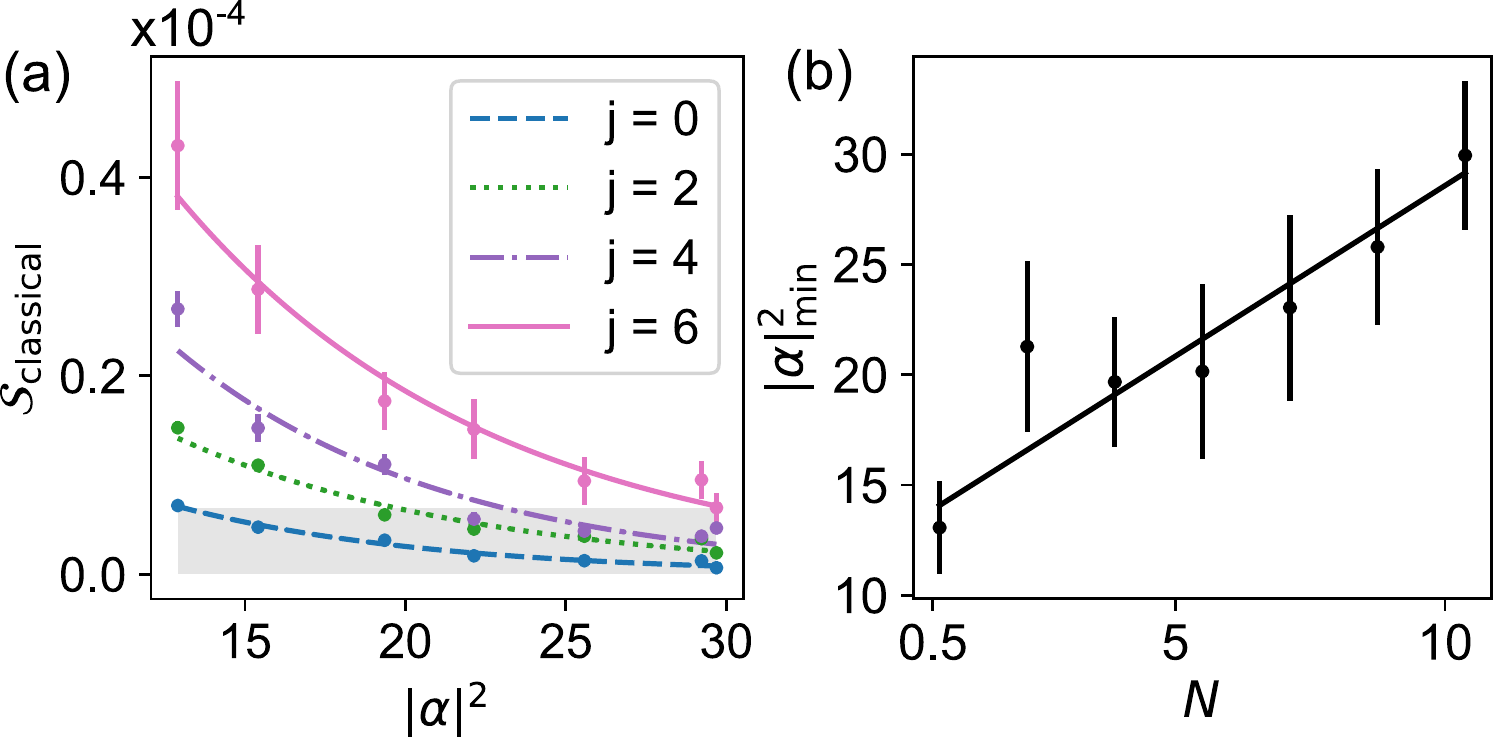}
	\caption{\textbf{Quantifying the transition towards a quadrature measurement.}
    	(a) We plot $\mathcal{S}_{\mathrm{classical}}$, the discrepancy between our measured data and an ideal quadrature measurement, as a function of $|\alpha|^2$ for various herald outcomes $j$. As $|\alpha|^2$ increases, our measurement becomes more quadrature-like and so $\mathcal{S}_{\mathrm{classical}}$ decreases. The gray box is the threshold used to define $|\alpha|^2_{\mathrm{min}}$. Error bars are one standard deviation in 10 trials.
    	(b) The minimum coherent state strength, $|\alpha|^2_{\mathrm{min}}$, required for a quadrature measurement.
    	We observe a linear scaling of $|\alpha|^2_{\mathrm{min}}$ with the average photon number in the signal, $N$. The error bars are obtained from the uncertainty in the fit parameters $A$ and $B$.}
	\label{fig:quantifying_transition}

\end{figure}

To understand why, we quantify the transition from $\Delta\hat{n}$ to $\Delta\hat{n}_\mathrm{classical}$ by computing the sum of the squared residuals,
\begin{equation}
\mathcal{S}_{\mathrm{classical}} = \frac{1}{\nu} \sum_{\Delta n} |P^{(j,\alpha)}_{\mathrm{exp}}(\Delta n) - P^{(j,\alpha)}_{\mathrm{classical}}(\Delta n)|^2,
\label{eqn:sumRes}
\end{equation}
where $\nu$ is the number of data points.
$\mathcal{S}_{\mathrm{classical}}$ quantifies the discrepancy between the measured data $P^{(j,\alpha)}_{\mathrm{exp}}$ and the classical model $P^{(j,\alpha)}_{\mathrm{classical}}(\Delta n)$.
Ideally, $\mathcal{S}_{\mathrm{classical}}$ smoothly converges to zero for increasing $|\alpha|^2$ as the validity of the classical field approximation improves.
In Fig.~\ref{fig:quantifying_transition}(a), we plot $\mathcal{S}_{\mathrm{classical}}$ for four different herald outcomes $j$.
We found heuristically that an exponential curve adequately models $\mathcal{S}_{\mathrm{classical}}$ for sufficiently large $|\alpha|^2$.
Thus, for each $j$, we fit $\mathcal{S}_{\mathrm{classical}}$ to $A\exp{\left(-B|\alpha|^2\right)}$ from which we determine $|\alpha|^2_{\mathrm{min}}$, the coherent state strength required to reach below $\mathcal{S}_{\mathrm{classical}} = 6.7\times10^{-6}$.
This threshold corresponds to the sum of the squared residuals obtained with the quantum model [i.e. replacing $P^{(j,\alpha)}_{\mathrm{classical}}(\Delta n)$ with $P^{(j,\alpha)}(\Delta n)$ in Eq.~\eqref{eqn:sumRes}], averaged over all $j$ and $|\alpha|$.
We plot $|\alpha|^2_{\mathrm{min}}$ for $j=0$ to $6$ in Fig.~\ref{fig:quantifying_transition}(b).
Interestingly, we observe a linear scaling between $|\alpha|^2_{\mathrm{min}}$ and the average photon-number in the signal, $N$, instead of a quadratic scaling~\cite{TS04,SDS17}.
We believe this relaxed requirement on $|\alpha|^2_{\mathrm{min}}$ is due to our detector inefficiency which smooths the fine features in the quadrature distribution of $\hat{\rho}^{(j)}_a$.
This reasoning agrees with the findings of Ref.~\cite{B90}, which showed that a smaller $|\alpha|^2_{\mathrm{min}}$ is required for states with smooth quadrature distributions such as coherent states.
Thus, the transition from photon-number to quadrature measurement in Fig.~\ref{fig:transitionPlot} occurred for a weaker $|\alpha|$ than might be expected (i.e. before the regime $|\alpha| \gg N$) since the quadrature distribution of $\hat{\rho}^{(6)}_a$ is smooth.

\begin{figure}[t!]
	\includegraphics[width=1\columnwidth]{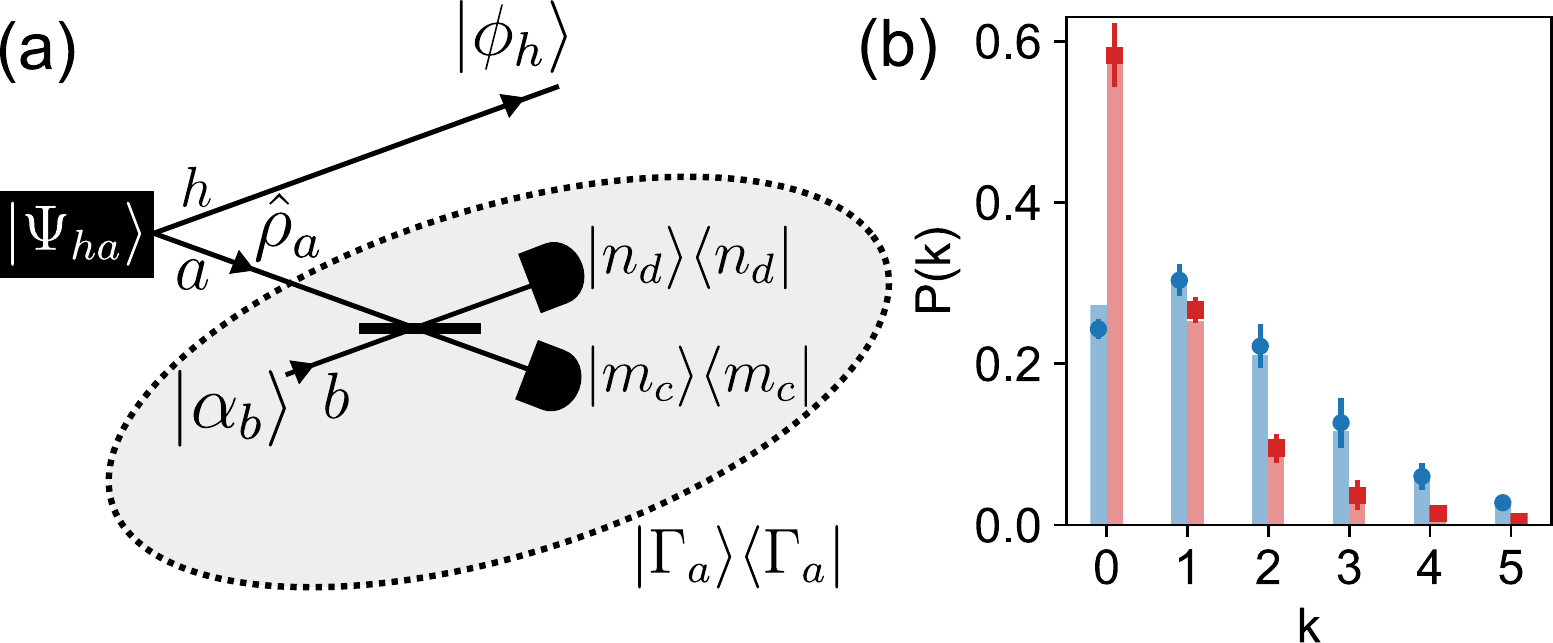}
	\caption{\textbf{State engineering using weak-field homodyne.}
	(a) A schematic of the concept. The weak-field homodyne detector, shown in the gray circle, projects the signal $\hat{\rho}_a$ onto the state $\ket{\Gamma_a}$ that depends on $m$, $n$, and $\alpha$. (b) Measured photon-number distribution of the state in the herald mode $P(k) = \left|\braket{k_h|\phi_h}\right|^2$ conditioned on obtaining the detection outcome $(m,n)=(6,0)$ when $|\alpha|^2 = 15.41$. The blue circles are measured when $\hat{\rho}_a$ and $\ket{\alpha_b}$ are temporally overlapped and thus interfere at the beam splitter. The red squares are measured when there is no temporal overlap between the two. Error bars are the standard deviation in 10 trials. The bars are theoretical predictions.}
	\label{fig:heraldedPn}
\end{figure}

\textit{State engineering}. So far, we have demonstrated that WFHD can tune between performing photon-number and quadrature measurements.
Here, we show it can also be used as a state engineering tool by projecting one part of a photon-number entangled state onto a particular measurement basis, thus steering the possible measurement outcomes on the other part of the entangled state.
The concept is shown schematically in Fig.~\ref{fig:heraldedPn}(a).
As before, mode $a$ of a two-mode squeezed vacuum state $\ket{\Psi_{ha}}$ is sent to the weak-field homodyne detector.
The detector projects mode $a$ onto a state $\ket{\Gamma_a}$ which depends on the detection outcome $(m,n)$ and $\alpha$~\cite{PLBC-RSW09}.
This measurement transforms mode $h$ to the state $\ket{\phi_h} = \mathcal{N}\braket{\Gamma_a|\Psi_{ha}}$, where $\mathcal{N}$ is a normalization factor. 
Many different classes of states $\ket{\phi_h}$ can be heralded since $\ket{\Gamma_a}$ can be continuously tuned between photon-number and quadrature-like states.
This versatility makes WFHD a powerful state engineering tool. 
For example, it was recently shown that $\ket{\phi_h}$ can have nearly perfect fidelity with a Schr\"odinger cat state $\ket{\beta} + \ket{-\beta}$ of arbitrary amplitude $\beta$~\cite{thekkadath2019engineering}.

As a proof of concept, we consider the specific detection outcome $(m,n) = (6,0)$ when $|\alpha|^2 = 15.4$.
The measured photon-number distribution $P(k) = \left|\braket{k_h|\phi_h}\right|^2$ is shown in Fig.~\ref{fig:heraldedPn}(b), which we use to calculate the second-order correlation function $g^{(2)} = \sum_k (k^2-k)P(k) / (\sum_k kP(k))^2$.
When $\hat{\rho}_a$ and $\ket{\alpha_b}$ are not temporally overlapped (using a temporal delay much smaller than the detection window), the two do not interfere and $P(k)$ is a thermal-like distribution (red squares, $g^{(2)} = 1.59 \pm 0.15$). 
This is because the average number of photons in mode $b$ (15.4) is much larger than in mode $a$ (1.74), and so the detection outcome provides little information about the number of photons in mode $a$.
In contrast, when $\hat{\rho}_a$ and $\ket{\alpha_b}$ are temporally overlapped, $P(k)$ changes drastically and resembles a Poisson distribution (blue circles, $g^{(2)} = 1.19 \pm 0.11$).
This demonstrates that the heralded state is strongly modified by the interference between $\hat{\rho}_a$ and $\ket{\alpha_b}$.

\textit{Conclusions}. We experimentally demonstrated that weak-field homodyne is a versatile detection scheme that can tune between photon-number and quadrature measurements.
We were able to demonstrate this tunability with quantum signals having up to nearly 11 photons. 
Our work shows that the dynamic range, efficiency, and noise-level of current photon-number-resolving detectors is enough to probe both particle and wave properties of quantum signals using weak-field homodyne detection. 

Another important achievement of our work is performing state engineering using weak-field homodyne for the first time. 
In particular, we showed that the photon-number distribution of a state heralded using weak-field homodyne can be strongly modified by the interference between the local oscillator and the signal. 
This paves the way towards promising applications in hybrid discrete- and continuous-variable quantum information processing protocols~\cite{AN-NVF15}. 
For example, weak field homodyne can be used to herald phase-sensitive non-Gaussian states such as two and four-component Schr\"odinger cat states~\cite{thekkadath2019engineering}.
An exciting prospect would be to verify the quantum non-Gaussian features of this heralded state using a second instance of the weak-field homodyne detector, which has the added advantage over regular homodyne detection of being robust against losses~\cite{R77,LCGS10,LSV15,HSRHMSS-S16,BTBSSV18,SPBTEWLNLG19}.
The same experimental setup could also be used to perform a hybrid Bell test~\cite{GPY88,BW98,KWM00,JKLZN10,DBJVDBW14}.

\textit{Acknowledgements}. The authors are grateful for discussions with J. Sperling, and thank J. Renema for his assistance with the installation of the cryogenic infrastructure. 
This work was supported by the following:  the Natural Sciences and Engineering Research Council of Canada (NSERC); the Networked Quantum Information Technologies Hub (NQIT) as part of the UK National Quantum Technologies Programme Grant EP/N509711/1); the EPSRC Programme Grant "Building Large Optical Quantum States" (EP/K034480); the Fondation Wiener Anspach; the European Commission project QUCHIP (H2020-FETPROACT-2014) and the ERC grant MOQUACINO (AdG).

\bibliographystyle{apsrev4-1}
\bibliography{refs.bib}



\newpage
\onecolumngrid

\input{./SM/sm.tex}


\end{document}

%% file: SM/sm.tex
\section*{Supplemental material}

\renewcommand{\theequation}{S\arabic{equation}}
\renewcommand{\thefigure}{S\arabic{figure}}

\subsection*{Theory I: Photon-number difference statistics without a classical field approximation}
\begin{figure}[ht]
	\includegraphics[width=0.3\columnwidth]{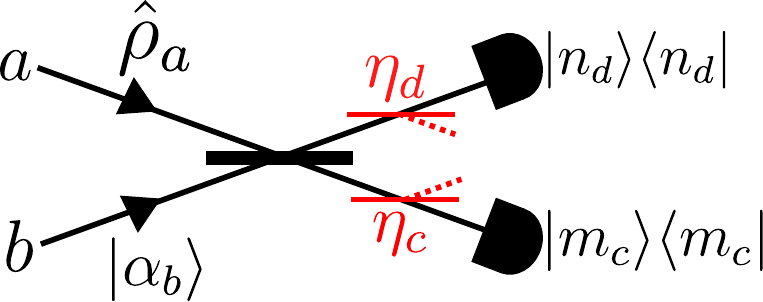}
	\caption{\textbf{Schematic of experiment.} 
	A signal $\hat{\rho}_a$ is combined with a coherent state $\ket{\alpha_b}$ on a balanced beam splitter.
	We measure the joint photon-number distribution at the output of the beam splitter.
	Detection efficiency is modelled by placing fictitious beam splitters of transmissivity $\eta_c$ and $\eta_d$ before the detectors in modes $c$ and $d$, respecitvely.
	}
	\label{fig:theory_schematic}
\end{figure}

Here, we derive the photon-number difference statistics measured by our detector without invoking a classical field approximation. 
The general setup is shown in Fig.~\ref{fig:theory_schematic}.
We begin by deriving the joint photon-number distribution measured at the output of the beam splitter in the idealized case where our signal is in a pure photon-number state  (and single spatio-temporal mode), $\hat{\rho}_a = \ket{j_a}\bra{j_a}$, and assume perfect detection efficiency, $\eta_c = \eta_d = 1$.
The joint probability of detecting $m$ photons in mode $c$ and $n$ photons in mode $d$ (i.e. the outcome $(m,n)$) at the output of the beam splitter, $P^{(j,\alpha)}(m,n)$, is given by:
\begin{equation}
\begin{split}
    P^{(j,\alpha)}(m,n) &= \left| \braket{m_c, n_d|j_a, \alpha_b} \right|^2 \\
    &=  \frac{1}{m!n!} \left| \braket{0_c, 0_d|\hat{c}^{m}\hat{d}^{n}|j_a, \alpha_b} \right|^2,
\end{split}
\label{eqn:joint_prob_ideal}
\end{equation}
where $\hat{c}$ ($\hat{d}$) is the photon annihilation operator in mode $c$ ($d$).
The input and output modes of the beam splitter are related through the following transformation:
\begin{equation}
\begin{split}
    \hat{c} = (\hat{a}+i\hat{b})/\sqrt{2}, \\
    \hat{d} = (\hat{b}+i\hat{a})/\sqrt{2},
\end{split}
\end{equation}
Inserting these expressions into Eq.~\eqref{eqn:joint_prob_ideal}, we find:
\begin{equation}
\begin{split}
P^{(j,\alpha)}(m,n) &= \frac{1}{2^{m+n}m!n!}\left| \braket{0_c, 0_d|(\hat{a}+i\hat{b})^m(\hat{b}+i\hat{a})^n|j_a, \alpha_b} \right|^2 \\
&= \frac{1}{2^{m+n}m!n!}\left| \braket{0_c, 0_d|\sum_{l=0}^m\sum_{k=0}^n{m \choose l}{n \choose k}i^{l+k}\hat{a}^{m-l+k}\hat{b}^{l+n-k}|j_a, \alpha_b} \right|^2.
\end{split}
\label{eqn:joint_prob_ideal2}
\end{equation}
The non-zero terms in the first sum satisfy $l = m+k-j$.
Furthermore, $\hat{b}\ket{\alpha_b} = \alpha\ket{\alpha_b}$.
Using these two facts, Eq.~\eqref{eqn:joint_prob_ideal2} can be simplified to:
\begin{equation}
P^{(j,\alpha)}(m,n) = \frac{e^{-|\alpha|^2}j!}{2^{m+n}m!n!}|\alpha|^{2(m+n-j)}\left| \sum_{k=0}^j{m \choose m+k-j}{n \choose k}(-1)^{k}\right|^2,
\label{eqn:joint_prob_ideal_final}
\end{equation}
and $P^{(j,\alpha)}(m,n)=0$ when $m+n < j$.


We now consider the effect of detection efficiency.
As shown in in Fig.~\ref{fig:theory_schematic}, we model detection efficiency by placing a fictitious beam splitter of transmissivity $\eta_c$ ($\eta_d$) before just before the detector in mode $c$ ($d$).
In this case, the joint probability to measure the outcome $(m,n)$ can be obtained by performing a Bernoulli transformation on Eq.~\eqref{eqn:joint_prob_ideal_final}~\cite{leonhardt1997measuring}:
\begin{equation}
P^{(j,\alpha)}_{\eta_c, \eta_d}(m,n) = \sum_{x=m}^{\infty}\sum_{y=n}^{\infty}{x \choose m}{y \choose n}\eta_c^{m}\eta_d^{n}(1-\eta_c)^{x-m}(1-\eta_d)^{y-n}P^{(j, \alpha)}(x,y).
\label{eqn:joint_prob_lossy_final}
\end{equation}
The intuition for this transformation is as follows. 
Suppose $m$ photons are detected in mode $c$ (we ignore mode $d$ to make the explanation easier to follow).
This detection event could have occurred when there were $x$ photons before the fictitious beam splitter, but $x-m$ photons were reflected, i.e. lost (where $x \geq m$).
Let $p(x)$ be the probability for there to have been $x$ photons before the beam splitter.
The probability to have transmitted $m$ photons and lost $x-m$ photons is $\eta_c^m$ and $(1-\eta_c)^{x-m}$, respectively. 
Since the detector does not discriminate between the photons, we must include the binomial factor ${x \choose m}$.
Thus, the total probability of detecting $m$ photons is the product of each of these probabilities, summed over all possible values of $x$, i.e. $\sum_{x=m}^{\infty} {x \choose m} \eta_c^m (1-\eta_c)^{x-m} p(x)$.
In Eq.~\eqref{eqn:joint_prob_lossy_final}, we apply a similar procedure but consider both modes $c$ and $d$.

We also consider the effect of mode mismatch (e.g. spatial, temporal, spectral, and polarization mismatch) between $\hat{\rho}_a$ and $\ket{\alpha_b}$.
In principle, both $\hat{\rho}_a$ and $\ket{\alpha_b}$ can occupy several spatio-temporal modes, thus making a full treatment of mode mismatch quite involved~\cite{RCCBS95}.
However, we found that the simpler approach of decomposing the problem into two effective orthogonal modes is sufficient to model our data.
This sort of heuristic model has been used and studied elsewhere, see e.g. Refs.~\cite{ourjoumtsev2006generating,tualle2009multimode}. 
Through a Gram-Schmidt process, $\ket{\alpha_b}$ can be decomposed in the following way \cite{T14}:
\begin{equation}
    \ket{\alpha_b} \rightarrow \ket{\sqrt{\mathcal{M}}\alpha_{||}}\otimes\ket{\sqrt{1-\mathcal{M}}\alpha_\perp},
\end{equation}
where $||$ denotes the same mode as the signal, $\perp$ denotes a mode orthogonal to the signal, and $\mathcal{M} \in [0,1]$ is a mode overlap parameter.
Since $\ket{\alpha_b}$ remains a coherent state when it is split across the modes $||$ and $\perp$, its amplitude is simply scaled by $\sqrt{\mathcal{M}}$ and $\sqrt{1-\mathcal{M}}$, respectively.
We wish to determine the joint probability of measuring the outcome $(m,n)$ when $\mathcal{M}\neq1$, i.e. $P^{(j,\alpha)}_{\eta_c, \eta_d, \mathcal{M}}(m,n)$.
We treat the detector as mode-insensitive, and so we convolve the joint probabilities of measuring the outcome $(m,n)$ in mode $||$ and mode $\perp$:
\begin{equation}
P^{(j,\alpha)}_{\eta_c, \eta_d, \mathcal{M}}(m,n) = P^{(j,\sqrt{\mathcal{M}}\alpha)}_{\eta_c, \eta_d}(m,n) * P^{(0,\sqrt{1-\mathcal{M}}\alpha)}_{\eta_c, \eta_d}(m,n),
\label{eqn:joint_prob_mode_mismatch}
\end{equation}
where $*$ denotes a convolution operation.
The term $ P^{(j,\sqrt{\mathcal{M}}\alpha)}_{\eta_c, \eta_d}(m,n)$ is the joint probability of measuring the outcome $(m,n)$ in the mode of the signal $||$.
Similarly, the term $P^{(0,\sqrt{1-\mathcal{M}}\alpha)}_{\eta_c, \eta_d}(m,n)$ is the joint probability of measuring the outcome $(m,n)$ in the mode orthogonal to the signal, i.e. $\perp$ (hence the reason why the signal is vacuum in this mode, i.e. $j=0$).

\begin{figure}[ht]
	\includegraphics[width=0.3\columnwidth]{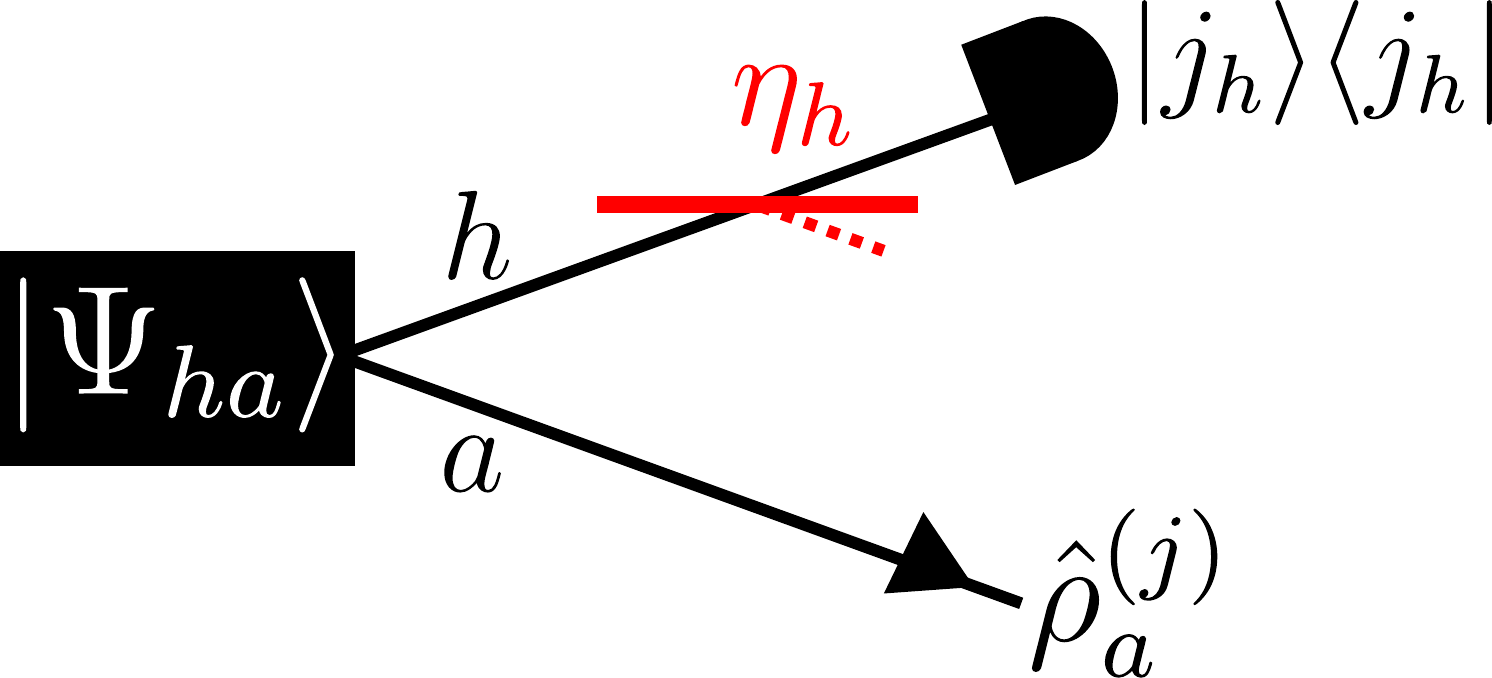}
	\caption{\textbf{Imperfect signal preparation.} 
    The imperfect signal $\hat{\rho}^{(j)}_a$ is prepared by performing a heralding measurement $\ket{j_h}\bra{j_h}$ on mode $h$ of a two-mode squeezed vacuum state $\ket{\Psi_{ha}}$.
    The detection efficiency in mode $h$ is $\eta_h$.
	}
	\label{fig:theory_schematic_2}
\end{figure}

Finally, we also consider the effect of an imperfect signal preparation.
Our signal $\hat{\rho}_a$ is prepared by performing a heralding measurement $\ket{j_h}\bra{j_h}$ on a two-mode squeezed vacuum state $\ket{\Psi_{ha}} = \sqrt{1-|\lambda|^2}\sum_{n=0}^{\infty} \lambda^{n}\ket{n_h,n_a}$, where $\lambda$ is a squeezing parameter (see Fig.~\ref{fig:theory_schematic_2}).
However, an imperfect detection efficiency $\eta_h$ transforms the heralding measurement to $\ket{j_h}\bra{j_h} \rightarrow \hat{\Pi}_h = \sum_{f=j}^{\infty} {f \choose j} \eta_h^{j}(1-\eta_h)^{f-j} \ket{f_h}\bra{f_h}$ (see text below Eq.~\eqref{eqn:joint_prob_lossy_final} for intuition)~\cite{leonhardt1997measuring}.
This imperfect heralding measurement $\hat{\Pi}_h$ prepares the signal $\hat{\rho}^{(j)}_a$ which is given by:
\begin{equation}
    \hat{\rho}_a^{(j)} = \mathcal{X} \mathrm{Tr}_h \left( \hat{\Pi}_h \ket{\Psi_{ha}}\bra{\Psi_{ha}} \right) = \mathcal{X} \sum_{f=j}^{\infty} {f \choose j} \eta_h^{j}(1-\eta_h)^{f-j} |\lambda|^{2f} \ket{f}\bra{f}.
    \label{eqn:input_state}
\end{equation}
where $\mathcal{X}$ is a normalization factor such that $\mathrm{Tr}_a\left(\hat{\rho}_a^{(j)}\right) = 1$. We wish to obtain the joint probability of measuring the outcome $(m,n)$ given the imperfect signal $\hat{\rho}_a^{(j)}$, i.e. $P^{(j,\alpha)}_{\eta_c, \eta_d, \eta_h, \mathcal{M}}(m,n)$.
Eq.~\eqref{eqn:input_state} is a statistical mixture of photon number states.
Thus, $P^{(j,\alpha)}_{\eta_c, \eta_d, \eta_h, \mathcal{M}}(m,n)$ can be found by computing $P^{(j,\alpha)}_{\eta_c, \eta_d,\mathcal{M}}(m,n)$ for each term in the mixture $\hat{\rho}_a^{(j)}$ and adding the probabilities with the appropriate weight, that is:
\begin{equation}
P^{(j,\alpha)}_{\eta_c, \eta_d, \eta_h, \mathcal{M}}(m,n) = \mathcal{X} \sum_{f=j}^{\infty} {f \choose j} \eta_h^{j}(1-\eta_h)^{f-j} |\lambda|^{2f} P^{(f,\alpha)}_{\eta_c, \eta_d,  \mathcal{M}}(m,n).
\label{eqn:joint_photon_final}
\end{equation}
The photon-number difference statistics, i.e. the probability to measure $\Delta n = m - n$ photons at the output of the beam splitter, can be obtained by performing the following sum:
\begin{equation}
\boxed{
    P^{(j,\alpha)}_{\eta_c, \eta_d, \eta_h, \mathcal{M}}(\Delta n) = \sum_{m= \mathrm{max}(0,\Delta n)}^{\infty} P^{(j,\alpha)}_{\eta_c, \eta_d, \eta_h, \mathcal{M}}(m,m-\Delta n).
    }
    \label{eqn:quant_photon_diff}
\end{equation}
Eq.~\eqref{eqn:quant_photon_diff} is used to calculate the red curves in Fig. 3.

\subsection*{Theory II: Photon-number difference statistics with a classical field approximation}

Here, we would like to invoke the classical field approximation $\hat{b} \rightarrow |\alpha|e^{i\theta}$ and obtain an expression analogous to $P^{(j,\alpha)}_{\eta_c, \eta_d, \eta_h, \mathcal{M}}(\Delta n)$.
We refer to reader again to Fig.~\ref{fig:theory_schematic}.
As before, we begin in the idealized case where $\hat{\rho}_a = \ket{j_a}\bra{j_a}$ and $\ket{\alpha_b}$ are combined on a balanced beam splitter.
For perfect detection efficiency, it was shown in Ref.~\cite{VG93} that the probability to measure $\Delta n$ photons at the output of the beam splitter after invoking the classical field approximation is giving by:
\begin{equation}
    p^{(j,\alpha)}(\Delta n) = \frac{1}{\sqrt{2\pi}|\alpha|} \frac{1}{2^j j!} \left | H_j\left( \frac{\Delta n}{\sqrt{2}\alpha}\right) e^{-\Delta n^2/ 4\alpha^2} \right | ^2,
    \label{eqn:classical_fock_homodyne}
\end{equation}
where $H_j$ is a Hermite polynomial of order $j$.
Notably, Eq.~\eqref{eqn:classical_fock_homodyne} is simply the quadrature distribution of a $j$ photon number state scaled by $\alpha$, i.e. $\Delta n \rightarrow \Delta n / \alpha$~\cite{leonhardt1997measuring}.
Eq.~\eqref{eqn:classical_fock_homodyne} is referred to as $P_{\mathrm{classical}}^{(j,\alpha)}(\Delta n)$ in the main text.

The effect of (imbalanced) detection efficiency was also considered in Ref.~\cite{VG93}:
\begin{equation}
    p^{(j,\alpha)}_{\eta_c, \eta_d}(\Delta n) =  p^{(j,g)}(\Delta \tilde{n}) * \frac{1}{\sqrt{2\pi}\sigma} e^{-\Delta \tilde{n}^2/2\sigma^2},
     \label{eqn:classical_fock_homodyne_wLoss}
\end{equation}
where * denotes a convolution, $g = \eta\alpha$, $\Delta\tilde{n} = \Delta n - \alpha^2(\eta_d-\eta_c)/2$, $\sigma = \sqrt{g^2 (1-\eta)/\eta}$, and $\eta = (\eta_c + \eta_d)/2$. 

We treat the effect of mode mismatch in the same way as in Eq.~\eqref{eqn:joint_prob_mode_mismatch}:
\begin{equation}
    p^{(j,\alpha)}_{\eta_c, \eta_d, \mathcal{M}}(\Delta n) = p^{(j,\sqrt{\mathcal{M}}\alpha)}_{\eta_c, \eta_d}(\Delta n) * p^{(0,\sqrt{1-\mathcal{M}}\alpha)}_{\eta_c, \eta_d}(\Delta n).
\end{equation}

Similarly, our imperfect signal preparation is treated in the same way as was done previously. Using Eq.~\eqref{eqn:input_state}, we obtain:
\begin{equation}
 \boxed{
    p^{(j,\alpha)}_{\eta_c, \eta_d, \eta_h, \mathcal{M}}(\Delta n) =\mathcal{X}\sum_{f=j}^{\infty} {f \choose j} \eta_h^{j}(1-\eta_h)^{f-j} |\lambda|^{2f}  p^{(f,\alpha)}_{\eta_c, \eta_d, \mathcal{M}}(\Delta n).
    }
    \label{eqn:classical_photon_diff}
\end{equation}
Eq.~\eqref{eqn:classical_photon_diff} is used to calculate the blue regions in Fig. 3 of the main text.



\subsection*{Theory III: State engineering using weak-field homodyne}

We can use Eq.~\eqref{eqn:joint_photon_final} to calculate the photon-number distribution of $\ket{\phi_h}$ since $P^{(j,\alpha)}_{\eta_c, \eta_d, \eta_h, \mathcal{M}}(m,n)$ is the joint probability to measure $j$ photons in mode $h$, $m$ photons in mode $c$, and $n$ photons in mode $d$.
That is, we compute $P^{(j,\alpha)}_{\eta_c, \eta_d, \eta_h, \mathcal{M}}(m,n)$ for every $j$ while fixing $(m,n)$ and $\alpha$. 
The resulting distribution is then normalized, thus yielding $P(j|m,n,\alpha) = \left|\braket{j|\phi_h}\right|^2$.
This strategy is used to calculate the red and blue bars in Fig. 5 of the main text.

\subsection*{Experimental details}
\label{app:exp_details}
\begin{figure}[ht]
	\includegraphics[width=0.5\columnwidth]{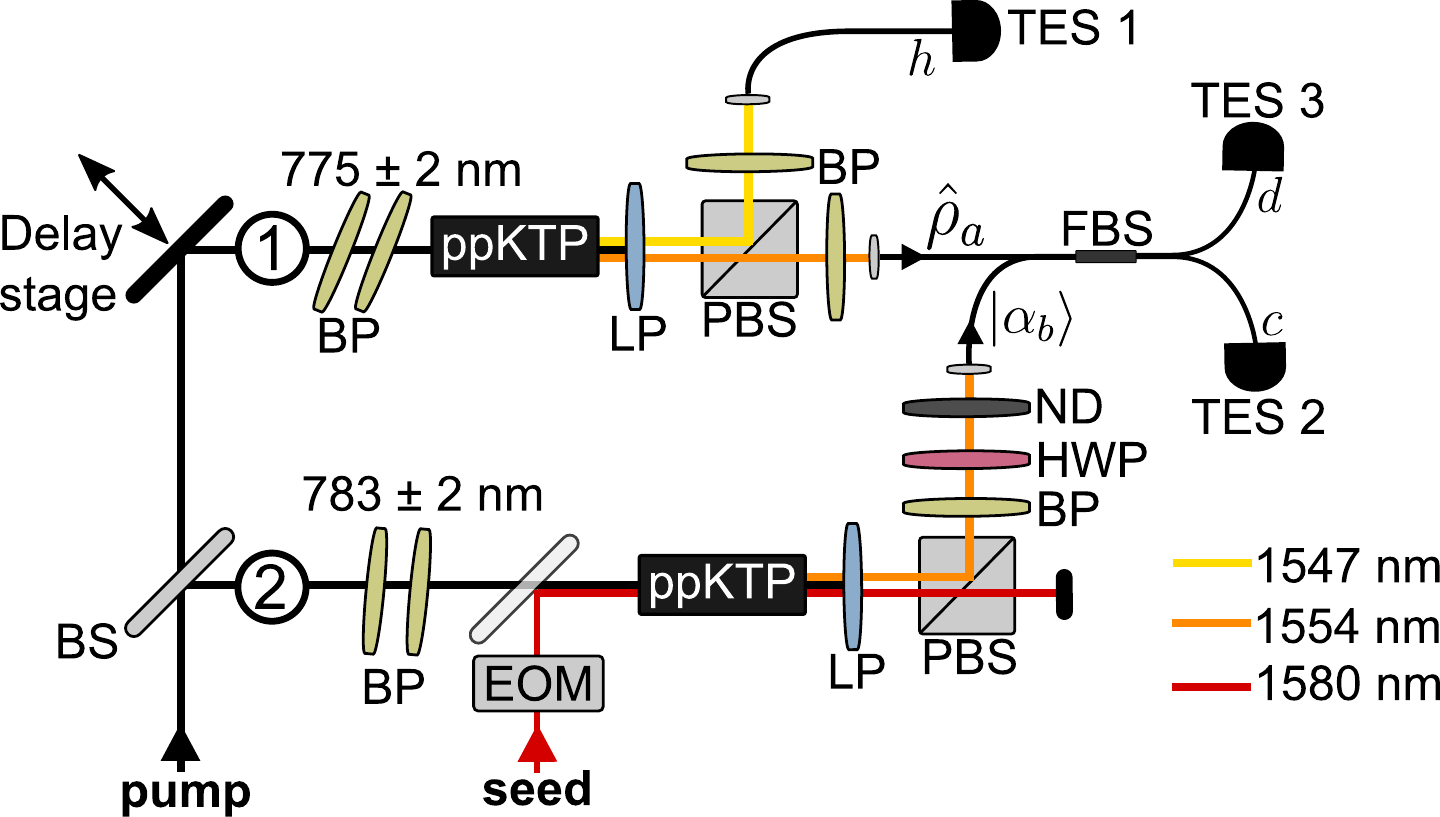}
	\caption{\textbf{Experimental setup.} Details can be found in the text.
	BS: beam splitter, BP: bandpass filter, EOM: electro-optic modulator, ppKTP: periodically-poled potassium titanyl phosphate (waveguide), LP: longpass filter, PBS: polarizing beam splitter, HWP: half-wave plate, ND: neutral-density filter, FBS: fiber beam splitter, TES: transition edge sensor.
	}
	\label{fig:exp:setup}
\end{figure}
The experimental setup is shown in Fig.~\ref{fig:exp:setup}.
The pump is a titanium sapphire oscillator followed by a regenerative amplifier that outputs femtosecond pulses (780 $\pm$ 20 nm [full width at half maximum]) at a rate of 100 kHz.
This rate is chosen to accommodate the thermal relaxation time of the transition edge sensor (TES) detectors.
The pump pulses are split into two paths, labelled \textcircled{\small{1}} and \textcircled{\small{2}} in Fig.~\ref{fig:exp:setup}.

In path \textcircled{\small{1}}, we prepare the signal $\hat{\rho}_a$.
The pump is filtered to 775 $\pm$ 2 nm using two angle-tuned bandpass (BP) filters.
The filtered pump is then coupled into a ppKTP waveguide and generates two-mode squeezed vacuum via type-II spontaneous parametric down-conversion.
The pump spectrum is chosen such that the down-converted signal (1554 nm) and idler (1547 nm) modes are approximately spectrally uncorrelated ($g^{(2)} = 1.849 \pm 0.007$ in idler mode).
The pump is then discarded with a longpass (LP) filter and the orthogonally-polarized signal and idler modes are spatially separated with a polarizing beam splitter (PBS).
Both modes are sent through a BP filter and coupled in fibers.
The idler mode is sent directly to a TES detector, whereas the signal mode is sent to a polarization-maintaining fiber beam splitter (FBS).

In path \textcircled{\small{2}}, we prepare the coherent state $\ket{\alpha_b}$.
The pump is filtered to 783 $\pm$ 2 nm and coupled into a second ppKTP waveguide.
In contrast to the previous path, we also couple 1580 nm light from a continuous-wave laser in order to seed the down-conversion process.
Through difference frequency generation, light with Poissonian photon-number statistics ($g^{(2)} = 1.005 \pm 0.002$, Fano factor $= 1.015 \pm 0.002$) is generated in the polarization mode orthogonal to the seed, i.e. $\ket{\alpha_b}$.
The seed and pump spectra are chosen to optimize the spectral overlap between the signal $\hat{\rho}_a$ and $\ket{\alpha_b}$ given that there are small differences between the phase matching properties of both waveguides.
$\ket{\alpha_b}$ is separated from the seed light using a PBS and a BP filter. To further minimize the amount of seed light leakage, we carve 2 ns pulses from the continuous-wave laser using an electro-optic modulator (EOM).
$\ket{\alpha_b}$ is sent through a half-wave plate (HWP) to match its polarization to that of the signal.
The intensity $|\alpha|^2$ is adjusted using a neutral-density (ND) filter wheel.
Finally, $\ket{\alpha_b}$ is coupled into fiber and sent to the FBS. 

\subsection*{Model parameters}

\begin{table}[h]
\centering
\begin{tabular}{|c  c|}
\hline
Parameter      		&  Value \\
\hline
$\eta_h$ 			& 0.395 $\pm$ 0.002 \\
$\eta_c$           	& 0.274 $\pm$ 0.001 \\
$\eta_d$            & 0.352 $\pm$ 0.002 \\
$|\lambda|$   		& 0.797 $\pm$ 0.001 \\
$\mathcal{M}$   	& 0.800 $\pm$ 0.060 \\
\hline
\end{tabular}
\label{table:params}
\caption{Measured parameters.}
\end{table}

We now describe how we determined the model parameters from various measurements.
Our model (see Eqs.~\eqref{eqn:joint_prob_mode_mismatch} and ~\eqref{eqn:classical_photon_diff}) requires the following five parameters: the detector efficiencies $\eta_c$, $\eta_d$, and $\eta_h$, the squeezing parameter $\lambda$, and the mode overlap parameter $\mathcal{M}$.
The results are summarised in Table I.

We determine $\mathcal{M}$ by measuring an interference signal. 
In theory, when the signal is a single photon, the probability to detect only one photon at both outputs of the FBS, $P^{(1,\alpha)}(1,1)$, vanishes.
In practice, this probability does not vanish due to a number of experimental imperfections, such as background counts, the modal purity of $\hat{\rho}_a$, and imperfect mode overlap between $\hat{\rho}_a$ and $\ket{\alpha_b}$.
As such, the visibility $\mathcal{V}$ of this interference signal provides a lower bound on $\mathcal{M}$.
We measured $\mathcal{V} = 0.800 \pm0.060$, and found that the models agree best with the data by using $\mathcal{M} = 0.82$.

The total system efficiencies are obtained via a Klyshko measurement~\cite{K80}.
For this measurement, we set $|\alpha|=0$ and decrease the pump power to ensure that $|\lambda| \ll 1$.
In this small squeezing limit, $\ket{\Psi_{ha}}$ consists mostly of photon pairs, i.e. higher order photon-number states can be neglected.
Then, $\eta_h$ is the probability that TES 1 detects a photon given that either TES 2 or TES 3 detected a photon.
Next, we determine the probability $\eta_s$ that either TES 2 or TES 3 detects a photon given that TES 1 detected a photon.
Because of the balanced beam splitter before TES 2 and TES 3, the measured $\eta_s$ is the the average of $\eta_c$ and $\eta_d$, i.e. $\eta_s = (\eta_c + \eta_d)/2$.
Moreoever, $R = \eta_c/\eta_d$ can be obtained by looking at the ratio of the number of detected photons at TES 2 and TES 3.
As such, we can obtain $\eta_d$ and $\eta_c$ from $\eta_d = 2\eta_s/(1+R)$ and $\eta_c = R\eta_d$, respectively.
Using this procedure, we find $\eta_h = 0.395 \pm 0.002$, $\eta_c = 0.274 \pm 0.001$, and $\eta_d = 0.352 \pm 0.002$.
Note that these are not intrinsic detector efficiencies as they include the efficiencies of fiber coupling, fiber transmission, and waveguide transmission.

Next, we determine the squeezing parameter $|\lambda|$.
To do so, we count the number of photons produced by $\ket{\Psi_{ha}}$ per pump pulse when $|\alpha|=0$.
We measure $\braket{\hat{n}_h} = 0.689 \pm 0.001$ photons at TES 1.
The number of photons before losses is $\braket{\hat{n}_h} / \eta_h$.
Thus, the squeezing parameter is given by $|\lambda| = \tanh[{\mathrm{arcsinh}(\sqrt{{\braket{\hat{n}_h} / \eta_h}}})] = 0.797 \pm 0.001 $.

Finally, $|\alpha|$ is measured by blocking the pump in path \textcircled{\small{1}} and counting the number of photons arriving at TES 2 and 3, i.e. $|\alpha| = \sqrt{\braket{\hat{n}_c}/\eta_c + \braket{\hat{n}_d}/\eta_d}$.

\subsection*{Transition edge sensors}

\begin{figure}[ht]
	\includegraphics[width=0.95\textwidth]{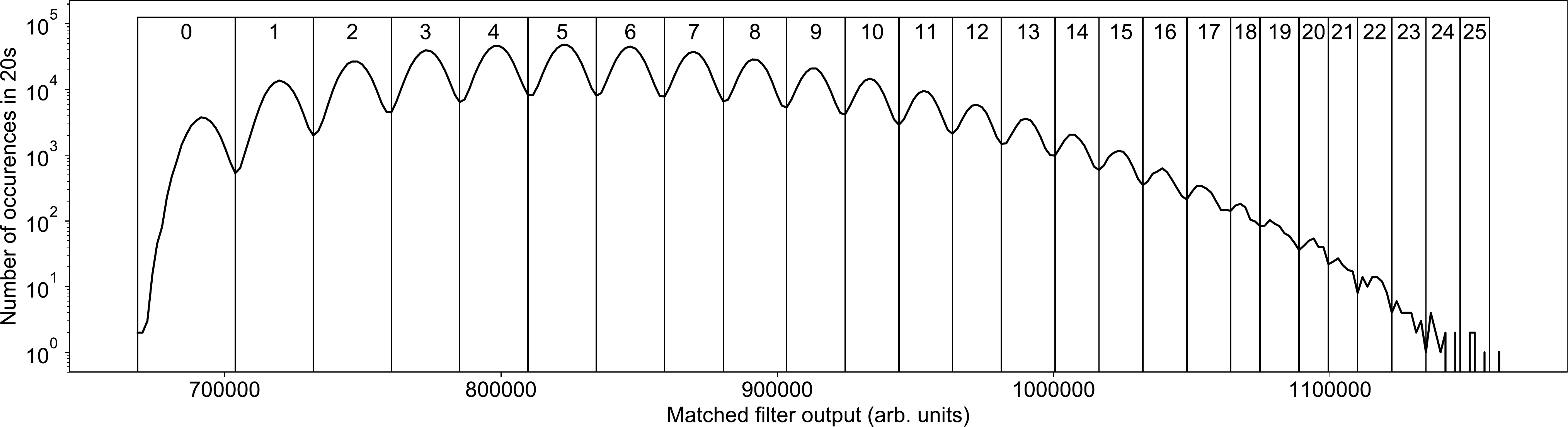}
	\caption{\textbf{Photon-number resolution of the transition edge sensor detector.} We plot a typical histogram produced by analysing the TES response following the procedure described in the main text. The corresponding photon-number is shown above each bin.}
	\label{fig:tes_hist}
\end{figure}

The transition edge sensors (TESs) operate in a dilution refrigerator at a temperature of around 80 mK.
Each TES consists of a superconducting tungsten film that is biased with a current in parallel with a 20 m$\Omega$ resistor such that the TES is near its transition to the non-superconducting state.
The resistor provides electrothermal feedback which stabilises the TES operating temperature~\cite{doi:10.1063/1.113674}.
When light is absorbed by the film, there is an increase in resistance that depends on the energy (and thus the number of photons) absorbed.
This generates an electrical response which is amplified using an array of superconducting quantum interference devices (SQUIDs) coupled to an inductor to provide cryogenic transimpedence gain~\cite{WM91}. 
This signal is then further amplified and filtered at room temperature.
The final electrical signal is read by an analogue-to-digital converter (ADC) triggered at the 100 kHz clock of the pump laser.
The trigger initiates a 10 $\mu$s acquisition on each TES.

Preliminary to data acquisition, we measure the average electrical response of the TES, which provides us with a matched filter~\cite{F-FCMPSW00}.
During the data acquisition, we calculate the overlap integral between this average trace and the detector response in real time.
The result of this overlap integral is a scalar value corresponding to an estimate of the energy absorbed by the TES.
As we only save the matched filter output, we minimize the amount of time spent writing data to the hard drive and thereby minimize the amount of trigger events lost due to data transfer bottlenecks.
It should be noted that this method of processing was chosen to maximise the data acquisition rate and that more sophisticated processing techniques exist~\cite{HMGHLNNDKW15, Levine:12}.

When we plot a histogram of the matched filter output, such as in Fig.~\ref{fig:tes_hist}, we see peaks due to the quantised energy of the optical field.
We use a peak finding routine to create bins for our data.
We then assign a label to each bin corresponding to its estimated photon number.

\subsection*{Testing the nonclassicality of our signal}
\label{app:test_NonCl}

\begin{figure}[ht]
	\includegraphics[width=0.5\columnwidth]{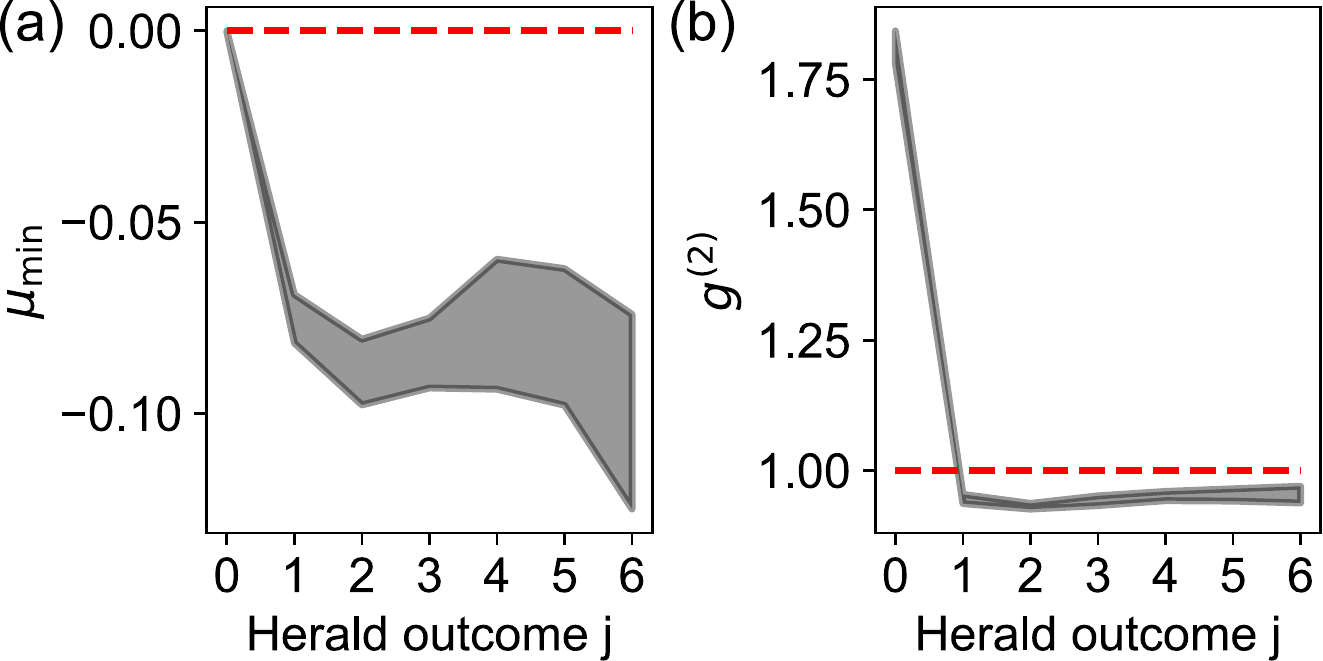}
	\caption{\textbf{Nonclassicality test.} We plot the results of the submultinomial (a) and sub-Poissonian (b) tests.
	Points lying below the red dashed line ($\mu_{\mathrm{min}} < 0$ and $g^{(2)} < 1$) have nonclassical statistics.
	The grey shaded region is calculated from experimental data, where the width corresponds to one standard deviation in 10 trials.}
	\label{fig:test_nonClass}
\end{figure}

Here, we test whether the photon-number statistics of our signal $\hat{\rho}^{(j)}_a$ have signatures of nonclassicality. 
We test for two different signatures, namely sub-Poissonian statistics~\cite{leonhardt1997measuring} and submultinomial statistics~\cite{sperling2017detector,sperling2017identification,SPBTEWLNLG19}.
For both tests, the coherent state is blocked, i.e. $\ket{\alpha_b} = \ket{0_b}$.
The results are shown in Fig.~\ref{fig:test_nonClass}.
We find that the state $\hat{\rho}^{(0)}_a$ has classical photon-number statistics and hence fails both non-classical tests, as expected.
However, for $j>0$, we measure both sub-Poissonian and submultinomial statistics with more than 1 standard deviation of confidence.
We now discuss how we performed each test.

To test for submultinomial statistics, we follow the procedure outlined in Refs.~\cite{sperling2017detector,sperling2017identification,SPBTEWLNLG19}. Namely, we determine the correlation matrix $M$ from our measured statistics. 
Let $E^{(j)}(k,l)$ denote the number of events where $j$ photons were detected at TES 1 (i.e. the herald TES), $k$ photons were detected at TES 2, and $l$ photons were detected at TES 3 for $j,k,l \in [0,6]$.
Also, let $E_{tot}^{(j)} = \sum_{k,l}E^{(j)}(k,l)$ denote the total number of events for a given herald outcome.
Following Eqs. (D1), (D2), and (D3) of Ref.~\cite{SPBTEWLNLG19}, the correlation matrix element $M^{(j)}_{x,y}$ for a particular herald outcome $j$ is given by:
\begin{equation}
\begin{split}
    M^{(j)}_{x,y} &= \frac{2}{E_{tot}^{(j)}} \sum_{k,l}\left( \delta_{k,x}\delta_{l,y} + \delta_{k,y}\delta_{l,x} \right) E^{(j)}(k,l) \\ 
    &+ \frac{1}{[E_{tot}^{(j)}]^2} \left(\sum_{k,l} (\delta_{k,x}+\delta_{l,x})E^{(j)}(k,l)\right)\\
    &\times\left(\sum_{k,l} (\delta_{k,y}+\delta_{l,y}) E^{(j)}(k,l) \right),
\end{split}
\label{eqn:corr_matrix}
\end{equation}
with $\delta$ denoting the Kronecker delta.
We compute $M^{(j)}_{x,y}$ for $x,y \in [0,6]$ then calculate the minimum eigenvalue $\mu_{\mathrm{min}}$ of the resulting matrix $M^{(j)}$. 
$\mu_{\mathrm{min}} < 0$ is a signature of submultinomial statistics and hence nonclassicality.

To test for sub-Poissonian statistics, we first determine the photon-number distribution of our signal $P^{(j)}(n) = \mathrm{Tr}(\ket{n_a}\bra{n_a}\hat{\rho}^{(j)}_a)$ from our measurements. 
The quantity $P^{(j)}(n)$ can be obtained by simply combining the statistics of TES 2 and TES 3, that is:
\begin{equation}
P^{(j)}(n) = \frac{1}{E_{tot}^{(j)}} \sum_{k+l=n} E^{(j)}(k, l).
\label{eqn:combined_stats}
\end{equation}
Using Eq.~\eqref{eqn:combined_stats}, we compute the second-order correlation function $g^{(2)}$ for each herald outcome $j$:
\begin{equation}
g^{(2)(j)} = \frac{\sum_n (n^2 - n) P^{(j)}(n)}{\left( \sum_n n P^{(j)}(n) \right)^2}.
\end{equation}
$g^{(2)} < 1$ is a signature of sub-Poissonian statistics and hence nonclassicality.
We note that, since the value of $g^{(2)(j)}$ is unchanged if loss is applied to $\hat{\rho}^{(j)}_a$, the procedure above is not affected by the efficiencies of TES 2 and TES 3.

